\documentclass[twocolumn,showpacs,floats,superscriptaddress,10pt]{revtex4}
\usepackage{graphicx}
\usepackage{amsmath}

\begin{document}

\title{Electronic fiber in graphene}
\author{Zhenhua Wu}
\email{zhwu@semi.ac.cn}
\affiliation{SKLSM, Institute of Semiconductors, Chinese Academy of Sciences,\\
P.O. Box 912, 100083, Beijing, China}

\begin{abstract}
We investigate theoretically the transmission properties through a p-n-p
junction on graphene. Here we show that the electronic transport property
presents deep analogies with light propagation. It originates from the
similarity between the linear spectrum of the Dirac fermions and photons
that obey the Maxwell's equations. We demonstrate the p-n-p channel acts as
an electronic fiber in which electrons propagate along the channel without
dissipation.
\end{abstract}

\maketitle
\small
In recent years, graphene, a single layer of carbon atoms arranged
in a hexagonal lattice, exhibits abundant physics and potential
applications.~\cite{Novoselov2,Zhang,Novoselov1,Geim,Neto} Quantum transport
properties in graphene have attracted increasing attentions both from the
fundamental physics and potential application in carbon-based
nano-electronics devices. The properties arise from the unique linear energy
dispersion and the chiral nature of electrons at the $K$ ($K^{^{\prime }}$)
point of the Brillouin zone, the energy spectrum exhibits a linear
dispersion that can be well described by the massless Dirac equation.~\cite{Neto} These massless
Dirac fermions, can be viewed as electrons that lose their rest mass~$m_{0}$%
, or as neutrinos that acquire the electron charge $e$. Electrons in
graphene are quite different from the electrons in conventional
semiconductor two dimensional electron gases.

\begin{figure}[b]
\centering
\includegraphics [width=0.9\columnwidth]{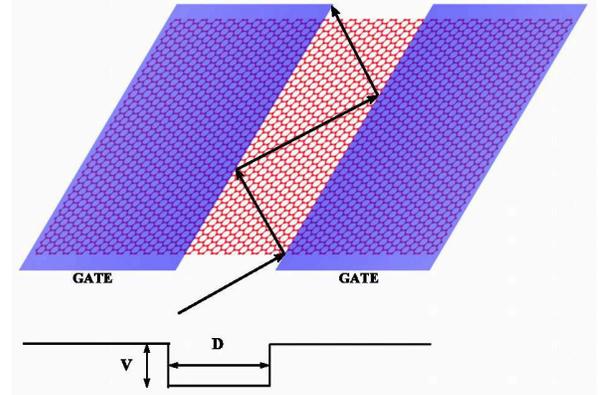}
\caption{\footnotesize Schematic of a p-n-p channel in graphene.
The shaded regions denote the regions below the electric gates. The lower
pannel describes the potential profile.}
\label{fig:model}
\end{figure}

Dirac equation describing the motion of massless quasiparticles is
mathematically similar to the Helmholz equation for an electromagnetic wave.
Graphene nano structures have been achieved by lithography as small as few
nanometers.~\cite{Ozyilmaz} While, the mean-free path of electron in
graphene approaches the order of micron at room temperature,~\cite{Geim} and
the electron wavelength is even lager.~\cite{Morozov} In such ballistic
regime, the scattering of electrons by potential barriers can be understood
by comparing with the reflection, refraction and transmission of
electromagnetic waves in inhomogeneous media. So, graphene monolayer is a
suitable candidate to examine the optical-like phenomena of the Dirac
fermions, that have attracted a lot of interest,~\cite%
{Cheianov,Jr,Park,FMZhang,Darancet,Beenakker,Ghosh,Zhao,Concha} such as waveguide, Goos-H$\ddot{a}$nchen effect,~\cite{Beenakker} Bragg
reflection, coherent buffers/memories, and
Brewster-type angle in graphene. Note that it was clearly
demonstrated by Cheianov et al.~\cite{Cheianov} that graphene based n-p-n
junctions could be used as electronic lenses or beam splitters.
In this letter, we investigated theoretically transport property of
Dirac electrons in a p-n-p channel created by two electrostatic gates that
are deposited on top of the graphene monolayer (see Fig.~\ref{fig:model}).
The transport process along the channel determined by the multiple total
internal reflection. We find that in such a p-n-p channel, electrons can be
confined in between two interfaces, in analogy with the light propagation in
an optical fiber. In addition, the approach presented in this work could
also be applied directly to electronic excitations in another recently
hotspot material, namely the topological insulator that also has Dirac
cones.~\cite{Qi,LBZhang}

\emph{Model---}Tunable potential barriers could be created on\ a
single-layer graphene sheet,~\cite{Huard,Williams} by electrostatic gates
deposited on top or back of the graphene monolayer. The electrostatic
potentials can be tuned by changing applied voltages (see Fig.~\ref%
{fig:model}). The low-energy electrons near the $K$ point of the Dirac cones
can be well described by the effective Hamiltonian~\cite{Wallace}
\begin{equation}
\hat{H}=\hbar v_{F}\boldsymbol{\sigma }\cdot \boldsymbol{k}+V^{i}(%
\boldsymbol{r},t),
\end{equation}%
where the superscript $i$ indicates the $i$-th region, $v_{F}$ is the Fermi
velocity, $\boldsymbol{\sigma }$ is the Pauli matrix and $V^{i}$ is the
height of the electrostatic barrier. The wavevector $\boldsymbol{k}$ satisfies $k_{x}^{i2}+k_{y}{}^{2}=(E-V^{i})^{2}/(\hbar v_{F})^{2}$. Note
that the translational invariance along the \textit{y} direction gives rise
to the conservation of $k_{y}$, and thus the solutions can be written as $%
\psi (x,y)=\psi (x)e^{ik_{y}y}$. For brevity, we will set $\hbar v_{F}\equiv
1$ in the follows.

\begin{figure}[tbp]
\centering
\includegraphics [width=0.9\columnwidth]{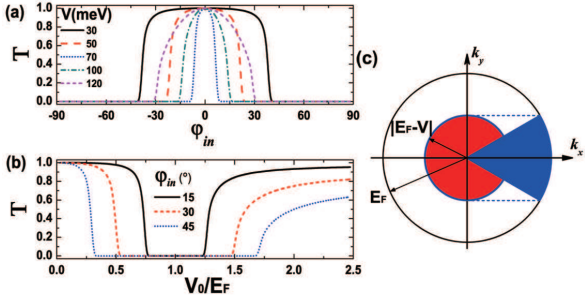}
\caption{\footnotesize (a) Transmission probability as a function
of the incident angle for several representative gate voltage V ranging from
30 meV to 120 meV. (b) Transmission probability as a function of the gate
voltage for different incident angles. The incident energy is fixed at $%
E_{F}=80$ meV both in panel (a) and (b). (c) The phase diagram in momentum
space. The outer circle indicates the wavevector in the incident n-doped
region and the inner circle indicates the wavevector in the transmitted
p-doped region.}
\label{fig:2}
\end{figure}

\emph{Refraction and total reflection---}We started by investigating
electron tunneling through an electric p-n junction on graphene, with $%
V^{in}=0$ in the n-doped region and $V^{out}=V$ in the p-doped region. This\
requirement fixes the dependence of the longitudinal outgoing wave vector $%
k_{x}^{\prime }=sgn(E_{F}-V)\sqrt{(E_{F}-V)^{2}-(k_{y})^{2}}$. Note that,
when $E_{F}-V<0$, $sgn(E_{F}-V)=-1$ denotes the interband scattering at the
interface of the p-n junction which is required by the helicity nature of
the Dirac Fermions. Continuity of the wave functions at $x=0$ gives the
transmission coefficient: $t=2k_{x}(E_{F}-V)/(E_{F}k_{x}+E_{F}k_{x}^{^{%
\prime }}-k_{x}V+ik_{y}V).$ Since the modes in n-doped and p-doped
regions have different group velocities, the transmission probability is
given by $T=\frac{v^{out}}{v^{in}}|t|^{2}$, where $v^{i}=\frac{1}{\hbar }%
\frac{\partial (E_{F}-V^{i})}{\partial k_{x}^{i}}=\frac{k_{x}^{i}}{%
E_{F}-V^{i}}$ for propagating states and $v^{i}=0$ for evanescent states.
This expression has the advantage that can well describe the propagating
states and evanescent states in a unified form. We can also define the
incident and refractive angles for electrons in propagating states $\varphi
_{in}\equiv \arctan (k_{y}/k_{x})$, and $\varphi _{out}\equiv \arctan
(k_{y}/k_{x}^{^{\prime }})$, and rewrite the transmission probability as
\begin{equation}
\footnotesize T=\frac{v^{out}}{v^{in}}\cdot \frac{4\text{cos}(\varphi
_{in})^{2}}{2+2sgn(E_{F}-V)\cos (\varphi _{out}+\varphi _{in}+\Theta \lbrack
-(E_{F}-V)]\cdot \pi )}.
\end{equation}%
This is valid for incident angle $\varphi _{in}<\varphi _{c}\equiv \arcsin
(|E_{F}-V|/E_{F})$, and the Heaviside step function $\Theta $ guarantees a
general expression of the transmission probability whether interband
scattering occurs or not. $\varphi _{c}$ is analogy to critical angle for
total reflection in optics and $\sin (\varphi _{out})/\sin (\varphi
_{in})=E_{F}/(E_{F}-V)\equiv n$ gives the Snell's law for transmitted
electrons. It shows that the effective Fermi energy ($E_{F}-V$) plays the
role of index of refraction in optical medium. Note that when $E_{F}-V<0$,
refractive index $n$ of graphene is negative as a metamaterial. In this
case, electrons are injected to the hole branch in \textit{p} region, the
chirality of Dirac electrons in graphene gives rise to sign reversal of the
momentum of the forward-going state. Importantly, one can tune the
refractive index $n$ in quite a large range via electrostatic gates, which
is not easy for normal metamaterials. We note en passant that such
optic-like behaviours are absent in the tunneling through a barrier,~\cite%
{Chen} since the effective Fermi energies at both side of the barrier are $%
E_{F}$, indicating an identical index of refraction.

The transmission spectra for electrons traversing such a p-n
junction are shown in Fig.~\ref{fig:2}(a) and (b), which can be understood
from the phase diagram of Fig.~\ref{fig:2}(c). For a fixed incident energy,
the transmission declines sharply and then is blocked when the incident
angle $\varphi _{in}$ exceeds a critical value $\varphi _{c}$ (see Fig.~\ref%
{fig:2}(a)), while the wavevector $k_{x}^{^{\prime }}$ in the p-doped region
become imaginary denoting the appearance of evanescent modes. The critical
angle $\varphi _{c}$ for total reflection can be estimated by the Snell's
law, $\varphi _{c}=\arcsin (|E_{F}-V|/E_{F})$. The transmission can also be
tuned by the gate voltage, or equivalently, the barrier height $V$ (see Fig.~%
\ref{fig:2}(b)). A transmission gap appears as the gate voltage $V$
approaching the incident energy $E_{F}$ and become wider as increasing the
incident angle $\varphi _{in}$. These tunneling features are well
appreciated by inspecting the phase diagram as shown in Fig.~\ref{fig:2}(c).
The Fermi wavevector is different in n-doped and p-doped region as a
consequence of gate voltage $V$, i.e., $k_{F,N}=E_{F}$, $k_{F,P}=|E_{F}-V|$,
while the component $k_{y}=k_{F,N}\sin (\varphi _{in})$ is conserved. The
requirement $k_{y}<k_{F,P}$ leads to non-zero transmission for allowed
incident angles as a sector in Fig.~\ref{fig:2}(c) in accordance with our
numerical results in Fig.~\ref{fig:2}(a). It is clear that $k_{F,P}$
declines as the gate voltage $V$ approaching the incident energy $E_{F}$, so
transmission gaps appear when $k_{F,P}$ drops below $k_{y}$ that verifies
the calculation as shown in Fig.~\ref{fig:2}(b).

\begin{figure}[h]
\centering
\includegraphics [width=0.9\columnwidth]{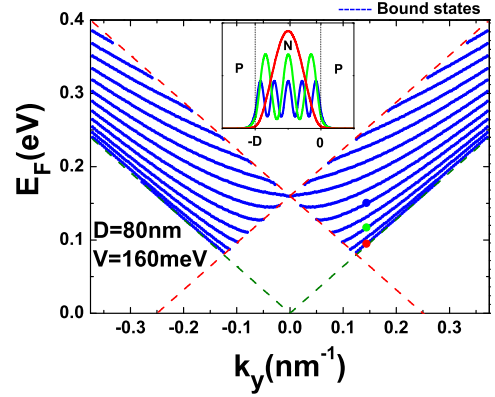}
\caption{\footnotesize Energy spectrum for the few lowest channel
modes for $V=160meV$, $D=80nm$. The inset shows the probability for states
indicated by solid circular dot with the same color.}
\label{fig:3}
\end{figure}

\begin{figure}[h]
\centering
\includegraphics [width=\columnwidth]{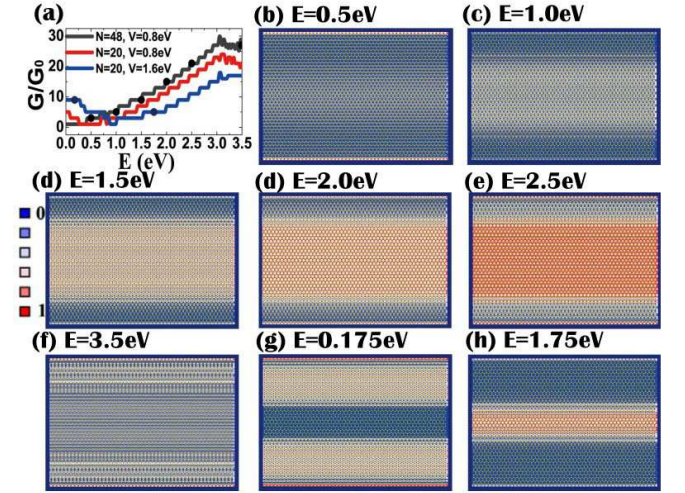}
\caption{\footnotesize (a) Conductance versus incident energy. $N$
labels the number of atoms contained in a cell of the middle channel.
(b)-(h) Density distribution of electron states corresponding solid circular
dot in Panel (a) for (b)-(f) $N=48$, $V=0.8$ eV, and (g)-(h) $N=20$, $V=1.6$
eV.}
\label{fig:4}
\end{figure}

\emph{Fiber guiding--}We considered an electric p-n-p channel with
width $D$ (see Fig.\ref{fig:model}) and focus on the electrons transporting
along the interface rather than across it. The transmission gap discussed
above can lead to confinement of electrons between the two interfaces
associated with multiple total internal reflections. Electrons can transport
along the p-n-p channel just like photons in an optical fiber. We calculated
the energy dispersion relation of the bound states in the case of total
internal reflection at both two interfaces (i.e. $\varphi _{in}>\varphi _{c}$%
, $k_{x}^{\prime }$ is imaginary). Matching of propagating waves to
evanescent waves at $x=-D$ and $x=0$, the energy spectrum of the bound
states can be obtained from the zero point of the determinant of the
coefficients, which can be reduced to a transcendental equation $%
-ik_{x}^{\prime }k_{x}\cos (k_{x}D)+[k_{y}^{2}-E_{F}(E_{F}-V)]\sin
(k_{x}D)=0.$

The energy spectrum $E_{F}(k_{y})$ is plotted for the first few
bound states in Fig.~\ref{fig:3}. The dashed lines $|k_{y}|=|E_{F}-V|$, and $%
|k_{y}|=|E_{F}|$ give the boundary that delimit the continuum region,
corresponding to propagating states in \textit{x}-direction. Confinement of
electrons in \textit{x}-direction gives rise to non-dispersive transporting
in \textit{y}-direction. The bound states between the\ two interfaces thus
play the roles of guide modes along the channel that can effectively convey
the charge carriers with group velocities indicated by the slope of the
dispersion relation. The spectrum is symmetric with respect to the
transverse wave vector $k_{y}=0$, guaranteed by the time reversal symmetry.
So the transport properties is isotropic for each incident terminal. The
inset of Fig.~\ref{fig:3} shows the probability density of the bound states
marked by the solid circular dot. The vertical dashed lines indicate the
interfaces of the p-n-p junction. It is clear that the solutions are
standing waves in the channel and decay exponentially in the barrier
regions. Note that the order of the bound states equals to the number of
peaks of standing waves in the channel, e.g. the 1st, 3rd and 5th bound
states or guide modes are shown in the inset of Fig.~\ref{fig:3}. Thus, the
bound states serve as guide modes in cavity. Finally, we focused on how the
fiber guiding features shown above are reflected in the measurable quantity,
the conductance $G$. We performed numerical simulations of electrical
conduction in a tight-binding model of a graphene zigzag ribbon covered by
electrodes. The conductance is evaluated using the recursive Green function
technique and Landauer-B\"{u}ttiker formula. The conductance of the gated
ribbon displays a step-like feature, which corresponds to the opening of the
new modes as the Fermi energy increases (see Fig.~\ref{fig:4}(a)). When $%
E_{F}$ is very small, there is no bound states in the channel, the
conductance is contributed by the edge states of the zigzag ribbon, since
our simulation are performed in a lattice scale. In such low incident energy
region, the conductance for a thinner channel may\ also depend on the former
hole-branch bulk states that are lifted up to the Fermi energy by the gate
voltage. A lager gate voltage can make this effect more pronounced.  When
the incident Fermi energy lays in an interval around the gate voltage,
electrons could be totally reflected with a small enough critical angle
since there exist a great difference between the refraction indexes at two
sides of the interface. In this case, most transmission modes are formed by
the bound states in the channel and desired high effective fiber guiding is
obtained. As the Fermi energy is increased continuously, the refraction
indexes of the channel and the gated region approach to each other
gradually, so the total reflection critical angle becomes very large.
Electrons can leak out of the channel easily, leading to decrease of guide
efficiency. This is because the transmission modes are mixture of both bound
states and bulk states out of the channel. Above expectations are
demonstrated by examining the electron density distributions as shown in
Fig.~\ref{fig:4}(b)-(h)). It is apparent that electrons are mainly
distributed in the channel when the incident Fermi energy is around the gate
voltage and thus the conductance depends on the guide modes by virtue of the
gate voltage and channel width. In addition, similar results could be
obtained when the gate voltage or incident energy are minus due to the
particle-hole symmetry.

In summary, We investigate theoretically the electronic quantum
transport properties in a p-n-p junction on graphene. Our results
demonstrate that the proposed structure on graphene presents deep analogies
with optic phenomena as refraction and total internal reflection. The
quantized dispersion relation of the confined electron states are obtained
in our calculation. We show that these bound states serve as guide modes and
can carry current along the channel with out radiation. The investigation
could be helpful to offer a functionality of graphene based electronic
fiber.

\end{document}